# Hot Diggity Dog: Simultaneous CO and Dust Modeling of the Most Luminous *WISE* Hot Dust-Obscured Galaxy Unveils Extreme Molecular Gas Excitation

Kevin C. Harrington,[1,2,3,4] Román Fernández Aranda,[5,6] Leindert Boogaard,[7] Axel Weiß,[8] Tanio Diaz Santos,[6,9] Manuel Aravena,[4,10] Roberto J. Assef,[4] Chao-Wei Tsai,[11,12,13] Peter Eisenhardt,[14] and Daniel Stern[14]

1 Joint ALMA Observatory, Alonso de Córdova 3107, Vitacura, Casilla 19001, Santiago de Chile, Chile
2 National Astronomical Observatory of Japan, Los Abedules 3085 Oficina 701, Vitacura 763 0414, Santiago, Chile
3 European Southern Observatory, Alonso de Córdova 3107, Vitacura, Casilla 19001, Santiago de Chile, Chile
4 Instituto de Estudios Astrofísicos, Facultad de Ingeniería y Ciencias, Universidad Diego Portales, Av. Ejército Libertador 441, Santiago, Chile
5 Department of Physics, University of Crete, 70013, Heraklion, Greece
6 Institute of Astrophysics, Foundation for Research and Technology–Hellas (FORTH), Heraklion 70013, Greece
7 Leiden Observatory, Leiden University, PO Box 9513, NL-2300 RA Leiden, the Netherlands
8 Max-Planck-Institut für Radioastronomie, Auf dem Hügel 69, 53121 Bonn, Germany
9 School of Sciences, European University Cyprus, Diogenes Street, Engomi 1516, Nicosia, Cyprus
10 Millenium Nucleus for Galaxies (MINGAL)
11 National Astronomical Observatories, Chinese Academy of Sciences, 20A Datun Road, Beijing 100101, China
12 Institute for Frontiers in Astronomy and Astrophysics, Beijing Normal University, Beijing 102206, China
13 School of Astronomy and Space Science, University of Chinese Academy of Sciences, Beijing 100049, China
14 Jet Propulsion Laboratory, California Institute of Technology, 4800 Oak Grove Drive, Pasadena, 91109, CA, USA

**ABSTRACT**

Hot dust-obscured galaxies (Hot DOGs), the most infrared (IR) luminous objects selected by the *WISE* all-sky mid-IR survey, have yielded a sample of intrinsically luminous quasars (QSOs) with obscured nuclear activity and hot dust temperatures. The molecular gas excitation properties have yet to be examined in detail under such extreme conditions. Here we study the most far-IR luminous *WISE* Hot DOG W2246–0526, focusing on ALMA observations of the central host galaxy. Multi-J CO transition measurements at J=2-1, 5-4, 7-6, 12-11, and 17-16 provide the first self-consistent modeling constraints on the molecular gas and dust properties of any *WISE* Hot DOG to date, providing a benchmark for future studies of dust-obscured QSOs. We implement a state-of-the-art TUrbulent Non-Equilibrium Radiative transfer model (TUNER) that simultaneously models both the line and dust continuum measurements. Due to a combination of high molecular gas densities and high gas kinetic temperatures, this extreme CO spectral line energy distribution seems to turnover around the CO(12-11) transition, likely making this among the most highly excited galaxies ever reported. The model infers a molecular gas mass $\sim 8 \times 10^{10}$ $M_\odot$, and we conclude that J≥3 CO line luminosities trace the bulk of the molecular gas mass for this extreme system unlike low-excitation conditions calibrated locally. W2246–0526 is a rapidly evolving system, with a high mean value of the molecular gas kinetic temperature versus dust temperature $T_k$ / $T_d$ ∼ 4.3. This may be due to the shocks and outflows injecting kinetic energy within the central kpc of this host that have previously been reported. These signs of highly excited molecular gas in W2246–0526 motivates obtaining well-sampled CO ladders in larger samples to better understand the conditions within these short-lived episodes associated with the most obscured supermassive black hole activity.

## 1. Introduction

Hot dust-obscured galaxies (Hot DOGs) are extremely luminous unlensed obscured quasars (QSOs) originally detected by NASA's Wide-field Infrared Survey Explorer (*WISE*; Wright et al. 2010). The most luminous Hot DOG, WISE J224607.6–052634.9 (W2246–0526 hereafter), is located at $z = 4.601$ (Díaz-Santos et al. 2021), and is the most luminous obscured galaxy currently known (Tsai et al. 2015). This Hot DOG and its environment have been studied intensively over the past decade (Zewdie et al. 2023, 2025). Powerful active galactic nuclei (AGN) in galaxies often reflect intense activity by the central supermassive black hole (SMBH), and this Hot DOG is dominated by a dust-obscured SMBH that is driving outflows ($> 10,000\,\mathrm{km\,s^{-1}}$), shocks, and intense ionisation and turbulence from its immediate vicinity to the outskirts of the host galaxy (Tsai et al. 2018; Díaz-Santos et al. 2016, 2018; Vayner et al. 2024, Liao et al. submitted). The interstellar medium (ISM) of the host galaxy has recently been characterized through far-IR fine-structure lines observed with the Atacama Large Millimeter Array (ALMA) (Fernández Aranda et al. 2024); however, a detailed investigation into the molecular gas excitation properties was previously lacking. A common approach to characterizing the cold-to-warm molecular gas in galaxies is using CO observations in combination with non-local thermodynamic equilibrium (LTE), large velocity gradient (LVG) models (e.g. van der Tak et al. 2007; Weiß et al. 2007), typically assuming a single or double component with uniform density and temperature. Additional techniques have built upon this approach by simultaneously using CO emission lines and continuum measurements from the rest-frame far-IR, as well as a turbulence-driven log-normal gas density distribution (Strandet et al. 2017; Boogaard et al. 2020; Harrington et al. 2021; Jarugula et al. 2021). Here we apply this modeling to W2246–0526. Since the discovery of *WISE*-selected Hot DOGs (Eisenhardt et al. 2012; Wu et al. 2012) some works have reported CO analyses of kinematic properties (Penney et al. 2020; Sun et al. 2024; Martin et al. 2024). However, there has not been a modeling analyses of the rise and fall of the CO ladder in any of the Hot DOGS to constrain the physical molecular gas and dust properties of these heavily dust-obscured systems.

The purpose of this work is twofold: a.) determine the far-IR properties via modeling of the CO and dust emission from the central host of the most well-studied Hot DOG, W2246–0526 while b.) demonstrating insights gained from applying the state-of-the-art turbulent non-Local Thermodynamic Equilibrium (LTE) radiative transfer modeling to such extreme objects. We present both novel and archival ALMA and Very Large Array (VLA) measurements in §2, our modeling results and discussion are presented in §3, with a summary in §4 that provides an outlook for future work to examine the rise and fall of the CO line emission and dust spectral energy distribution (SED) in such extreme systems. We adopt a ΛCDM cosmology: $H_0 = 70$ km s$^{-1}$ Mpc$^{-1}$, $\Omega_m = 0.3$ and $\Omega_\Lambda = 0.7$. This corresponds to 6.536 kpc/″ at the source redshift, $z = 4.601$.

## 2. VLA CO(2-1) and ALMA Band 3-10 Measurements

Here we present new measurements of the ALMA observed CO(12-11) and CO(17-16) emission lines. The CO(17-16) line is only partially detected (see Appendix A), yet in the modeling we include a conservative 100% uncertainty in the extrapolated estimate for the total line flux as this line may also contain emission from OH+ (Gallerani et al. 2014). We also include previously reported CO(5-4) and CO(7-6) (Martin et al. 2024; Fernández Aranda et al. 2024). We use CO(2-1) line measurements from the VLA, combining the B and D configurations (Program ID: VLA/15B-192;VLA/17B-312) – the D configuration data was first presented in Díaz-Santos et al. (2018). We follow the methodology of Fernández Aranda et al. (2024, 2025), which had performed a resolved dust SED analyses. We extract the line and continuum emission from the central host (dot-dashed white circle, Figure 1), which has been separated with respect to the known companion galaxies in dust continuum maps (see also Fig. 1 of Fernández Aranda et al. 2025). The CO(2-1) line emission extends beyond the host galaxy, although at a lower signal-to-noise ratio (S/N). Tsai et al. (2018) derived a bolometric luminosity of W2246–0526 to be ~ 4 $\times 10^{14}$L$_\odot$ with up to ~ 2 $\times 10^{14}$ L$_\odot$ from both the mid-IR and the far-IR energy ranges.

The purpose of this work is to focus on the joint analyses of the dust emission and CO excitation from the central region of this dust-obscured QSO host, as defined in Fig. 1 and Fernández Aranda et al. (2025). The higher-J CO line measurements have a heterogenous mix of integration times (non-uniform sensitivity beyond the targeted emission from the host). To model the full extent of the molecular gas in and around this system, in a resolved fashion, is beyond the scope of this work. Comparably sensitive multi-J CO line measurements surrounding the central host would be required to determine a more complete estimate of the molecular gas mass and other properties, and the following physical properties of the cold ISM we derive are relevant to the emission only within the common aperture centered on the host. Appendix A presents more details on the data reduction, observational details, continuum-subtracted spectra and velocity-integrated line flux densities for each line (Fernández Aranda et al. 2024, 2025).

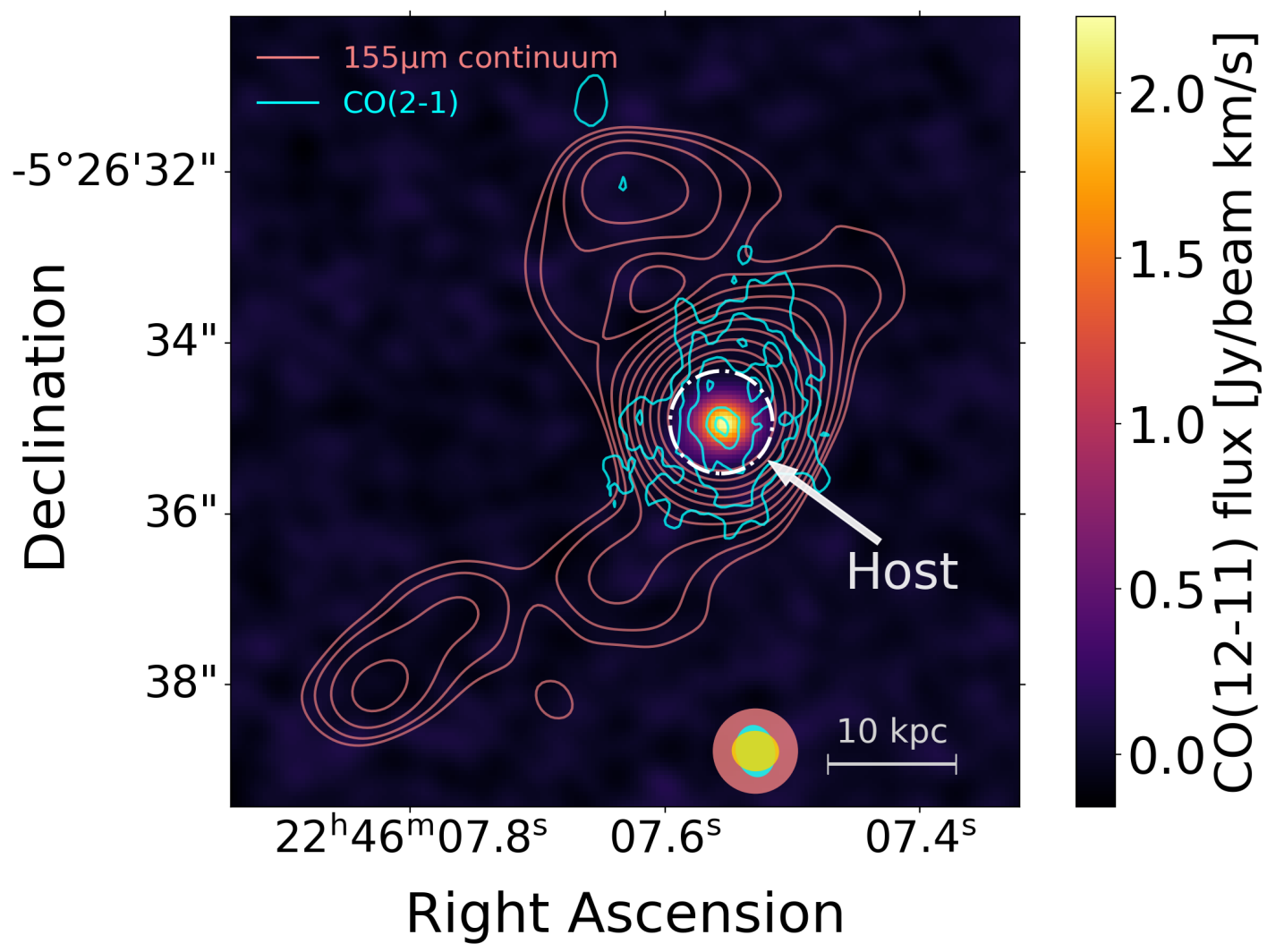

**Fig. 1.** The velocity-integrated CO(12-11) line emission from the central region of W2246–0526 (host), compared with extended dust continuum emission at rest-frame 155$\mu$m from Fernández Aranda et al. (2025) (pink contours), and the VLA B and D configuration CO(2-1) emission (cyan contours). The physical scale is shown in the lower right corner, with the corresponding synthesized beams (pink: dust continuum; cyan: CO(2-1); gold: CO(12-11)). The dot-dashed white circle indicates the r=0.6″aperture used to extract the flux density measurements of the host; i.e. not including any of the extended emission. Contours reflect [4,5,6,...] x r.m.s) for CO(2-1), and [3, $2^{(n/2)}$] x r.m.s. (with n=4,5,6,...) for the continuum.

## 3. TUNER Modeling of the molecular gas and dust

We simultaneously model the observed dust continuum and CO line fluxes using the TUrbulent Non-Equilibrium Radiative transfer (TUNER) model. The underlying framework for the TUNER model is described in detail in Harrington et al. (2021), with a few updates that will be described in detail in a forthcoming paper (Boogaard et al. in prep.). We also refer the reader to Strandet et al. (2017); Jarugula et al. (2021) for additional applications of the model. Briefly, the LVG approximations to the non-LTE radiative transfer calculations assume a lognormal volume density distribution (Padoan et al. 1997; Krumholz & McKee 2005) with a central density and width $\Delta V$ set by the turbulent velocity dispersion of the system. The different density bins that sample this distribution and are summed to derive the total observed emission and subsequent molecular gas and dust properties after optimizing for the free parameters that are described in Appendix §B) and listed in Table B.

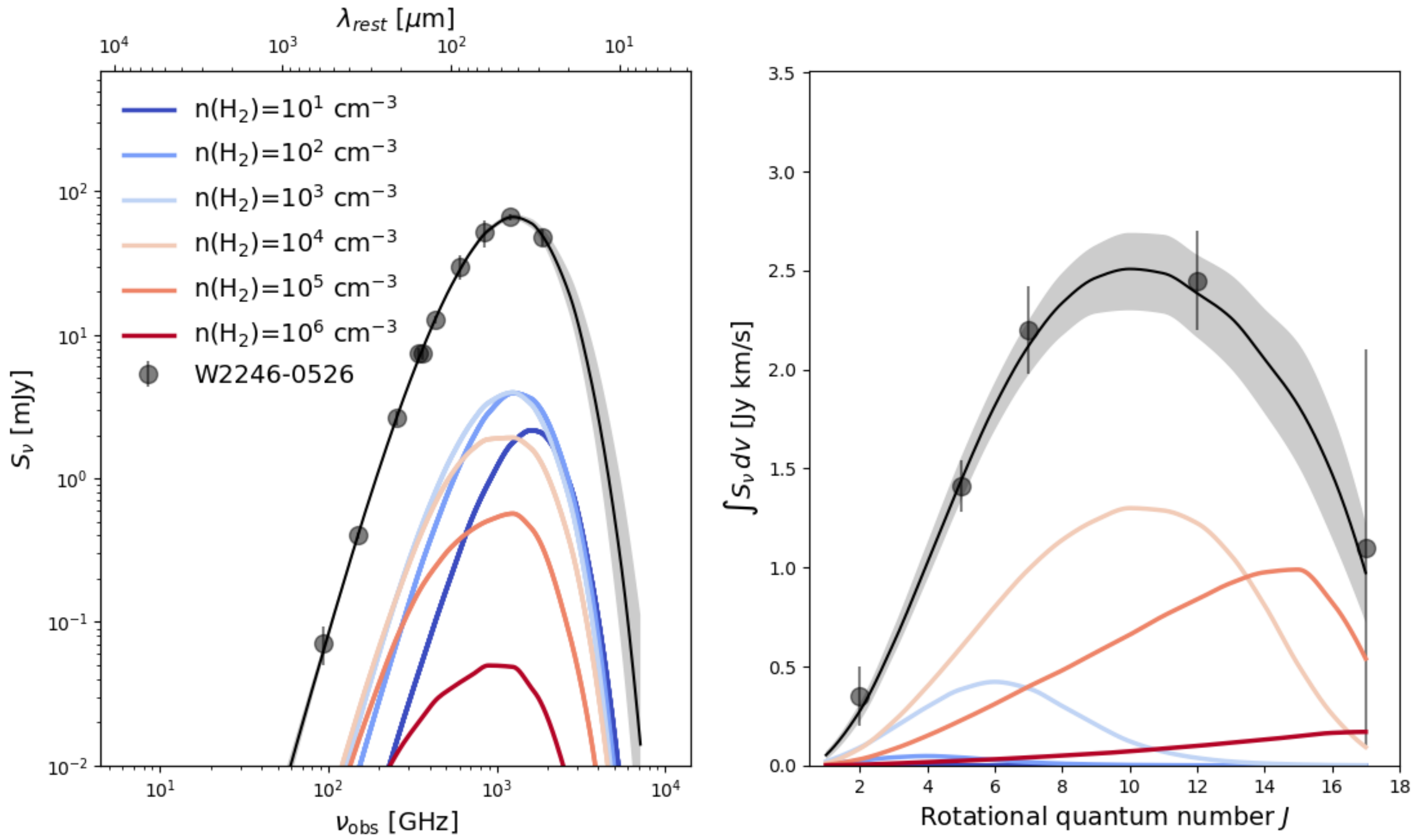

**Fig. 2. Left:** TUNER model fits to the observed data (black circles) for the dust and CO line SEDs. The 50th percentile is shown by a dotted black line and the 16-84th percentile range as a grey shaded region. The representative density contributions to the observed dust (**Left**) and CO line (**Right**) SED from each molecular gas density bin of log(n(H$_2$)) [cm$^{-3}$] = 1 (dark blue), 2 (blue), 3 (light blue), 4 (light orange), 5 (orange) and, 6 (red) cm$^3$ are shown in solid colored curves.

### *3.1. Line and Dust Spectral Energy Distribution*

Figure 2 reveals the CO line excitation ladder of W2246–0526, one of the most extreme CO ladders ever reported. Measurements between CO(7-6) and CO(12-11) prevent determining the exact peak in the CO line SED. The dispersion between the 16th and 84th percentile values in the model fits suggest such strong CO emission is present such that the turnover appears to be traced by the CO(12-11) transition in the central host of W2246–0526. Figure B indicates that the total CO emitting size area is $\sim$ 3 kpc$^2$ from the model-derived emitting radius of ~1 kpc. This is consistent with the effective stellar radius of the galaxy (Fan et al. 2018), and with the region from which we expect the CO and dust emission to arise from (Fernández Aranda et al. 2024, 2025). Although this result matches previous independent estimates, the finest spatial resolution corresponds to ~2.5 kpc. The CASA task `imfit` is used to fit a 2D Gaussian model of the CO(12-11) line emission, which is only marginally resolved. The deconvoled FWHM major and minor axis sizes of $1.7(\pm0.26)$ kpc $\times$ $0.98(\pm0.52)$ kpc provide an upper limit to the size, but it is consistent with the above results from the TUNER modeling and the effective stellar radius. We also note that Harrington et al. (2021) explains that an assumption in such a model is that the observed emission comes from a filled, face-on disk with a total source solid angle and effective radius. This total area is then divided into corresponding areas following the density distribution. This may not reflect the true source size, as it depends on how the molecular gas and dust emission is distributed within the source solid angle. The radius derived for each density therefore represents the typical sizes of those clouds, but not the actual size centered on the source. The dust SED is dominated by dust emission associated with molecular gas with density $\log(n(\mathrm{H_2})) \sim$ 2-4 cm$^{-3}$, a median dust temperature of ~100 K and $\beta_d \sim 2.3$. The median dust temperature is consistent with the uniform temperature derived in the modified blackbody fits of Fernández Aranda et al. (2025). In the TUNER modeling approach we did allow for possible values out to ~2000 K to consider solutions with warmer dust, although the emission that is described by the model corresponds only far-IR radiation. Here we neglect any shorter wavelength measurements of a possible hot dust component that may reveal an additional dust mass / heating source associated with this system and may further contribute to the observed line emission. This may be particularly important for explaining the excitation of the highest-J lines observed. Future work will be able to use the complete IR SED to better characterize this extreme system.

The CO emission lines have roughly equal contributions from densities $\log(n(\mathrm{H_2})) \sim$ 3-5 cm$^{-3}$ up to about J=3. We determine that a significant contribution to the observed high-J line emission arises from molecular gas with density $\log(n(\mathrm{H_2}))$ $\sim$ 4-5 cm$^{-3}$. Two examples of QSO systems with CO(12-11) or higher-J CO measurements include the lensed radio-quiet QSO BRI0952 (Omont et al. 1996; Kade et al. 2024) and the optically-selected QSO SDSSJ1148 (Fan et al. 2003; Gallerani et al. 2014), the latter with CO(17-16) detected, but no measurements between CO(7-6) and CO(17-16). The extreme CO excitation in W2246–0526 results in a CO(7-6) velocity integrated line intensity that is about 2-3.5 times higher than these highly excited QSO systems. The steeply rising CO line SED shape has previously been found in only a few systems, such as the strongly lensed, broad absorption line QSO APM 08279+5255 (magnified IR luminosity $\sim 7 \times 10^{15}$ $L_\odot$ hereafter 'APM0827') with measurements out to CO(11-10) (Irwin et al. 1998; Weiß et al. 2007). APM0827 has a highly uncertain magnification factor, reflecting the challenges in studying lensed objects. It still remains one of the few examples of the most highly excited CO emission that can be compared to the unlensed W2246–0526. Similar LVG modeling has been applied using a 2-component LVG model (Weiß et al. 2007) that is effectively the same underlying model as TUNER. The density distribution has been implemented in TUNER, which is the biggest change from the LVG model by Weiß et al. (2007). The overall parameter fits to the observed line SED for APM0827 provide a useful reference, however model uncertainties may have systematic differences that remain beyond the scope of this work (see, e.g., comparisons between 1- / 2-component models vs. those following density distributions in Harrington et al. 2021). Future applications of the TUNER model will examine the nature of CO line excitation in a larger sample of QSOs. APM0827 has a turnover at CO(10-9), and the high-J emission can be described by a dense component with $\log(n(H_2)) \sim 5$ $cm^{-3}$ within an apparent, magnified emitting size radius of ~900 pc. The intrinsic value is subject to the systematic uncertainty in the lens model, but is within 10s to 100s pc. W2246–0526 has similar densities and high excitation as APM0827, although the turnover in the CO line SED is around CO(12-11)! Lower densities cannot reproduce the observed line SEDs for both APM0827 and W2246–0526. The biggest difference (see Appendix Figure B.1) is found in the median values of $T_k$ at these densities for W2246–0526. The values of $T_k$ >300 K higher than the 60-70 K reported in APM0827 (Weiß et al. 2007). Therefore, W2246–0526 likely features the most highly excited global CO ladder reported in a galaxy to date.

The normalized CO partition function is shown in Figure 3. This demonstrates the contrast between the Milky Way (MW) center (Fixsen et al. 1999) and W2246–0526. In the MW, ~ 50% of the CO column resides in the J = 0 and J = 1 states, while this fraction drops to ~10% for W2246–0526. The bulk of the CO molecules exist at the lowest level populations in the Milky Way, reflecting the more extreme CO excitation in W2246–0526 as it has a broader distribution of level populations out to higher-J levels.

### 3.2. Characterizing the extreme ISM properties

W2246–0526 is known to have one of the highest reported bolometric luminosities of any galaxy. Our best-fit parameters derive a median IR luminosity $L_{IR:8-1000\mu m} = (2.41^{+0.11}_{-0.09}) \times 10^{13}$ $L_\odot$. Combined with the existence of an active SMBH with estimated mass $M_{SMBH} \sim 6 \times 10^9$ $M_\odot$(Tsai et al. 2018, Liao et al. submitted), one would suspect that this is the powering source for the higher-J line emission. From the best-fit line SEDs we determine that the molecular gas that contributes to the higher-J lines, with densities $\log(n(H_2)) \sim$ 4-5 $cm^{-3}$ (and gas kinetic temperatures $T_k \sim 250 - 500$K) subtends an area less than 100 pc on average for these dense clouds, yet higher spatial resolution is needed to determine the actual distribution within the source solid angle observed/modeled. One might assume that such compact regions responsible for the higher-J CO excitation are in agreement with the known dust-obscured SMBH in this central host, however the TUNER model does not specifically attribute the observed emission to any heating source nor any radial distribution. One can only speculate that such activity from the active SMBH of the central Hot DOG may be directly responsible for the extreme CO excitation. The high mean gas kinetic temperature $T_k \sim 408$ K, high dust temperatures $T_d \sim 96$ K, and importantly the high mean value of $< T_k / T_d > = 4.3 \pm 1.7$, all point towards the need for a strong powering source. Current spatial resolution prevents directly connecting the relatively high value of $< T_k / T_d >$ in the central core of this host Hot DOG to SMBH activity. This motivates future work with, e.g., the most extended configurations of ALMA. Harrington et al. (2021) alluded to this ratio as a potential indicator of the relative mechanical versus photoelectric heating of an object, therefore with a priori knowledge of this being a dust-obscured QSO, it is important to explore the known properties with respect to this high value. Independent estimates of the value of $T_k / T_d \sim 2$ have also been reported for local starbursts with $L_{IR} \sim 10^{11-12} L_\odot$ (Díaz-Santos et al. 2017). The fact that this ratio is close to four suggests strong mechanical energy within W2246–0526 as compared to starburst activity. The energetic activity from this obscured SMBH must therefore be able to permeate the ~3 $kpc^2$ emitting area to boost the $T_k / T_d$ ratio to high values. This is reflected by the TUNER model results with turbulent velocity dispersions of 100s $km\,s^{-1}$ that are consistent with reported FWHM measurements for W2246–0526, e.g. ~500 $km\,s^{-1}$ (Díaz-Santos et al. 2016, 2021). The high value of $T_k / T_d$ is also consistent with recent evidence of strong outflows and widespread shocks within the same area of W2246–0526 based on JWST spectra (Vayner et al. 2024). The turbulent gas in the PASSAGES (Harrington et al. 2016; Berman et al. 2022)[1] galaxies, with $L_{IR} \sim 10^{13-14}$ $L_\odot$, revealed relatively high values in the range between $T_k / T_d$ = 2.5-3.5, yet W2246–0526 has an even higher mean $< T_k / T_d > \sim 4.3$. This motivates further analyses of a wider population range (e.g. starbursts, AGNs, QSOs and Hot DOGs), with well-sampled dust and CO line SEDs, to examine whether or not $T_k / T_d$ values may be used as a diagnostic for the most extreme phases of galaxy evolution. One of the free parameters in the model that allows us to better interpret the high molecular gas kinetic-to-dust temperature nature of this source is $\gamma_T$, which helps to constrain the temperature-density degeneracy (see Fig. B.1), and is defined as a power-law slope index as $T_k \propto n^{\gamma_T}$ (Harrington et al. 2021). The posterior distributions suggest reasonable fits within the 16th and 84th percentiles of $\gamma_T$ ranging from ∼ -0.1 to 0.1. Typical molecular clouds have a negative $\gamma_T$ values, such that denser gas becomes colder (leading to collapse). However, in this extreme object the dust and CO line SEDs can be reproduced with a positive value of $\gamma_T$, implying dense gas is hotter than less-dense gas. This is independent of any knowledge that this is an active dusty SMBH with a strong central heating source.

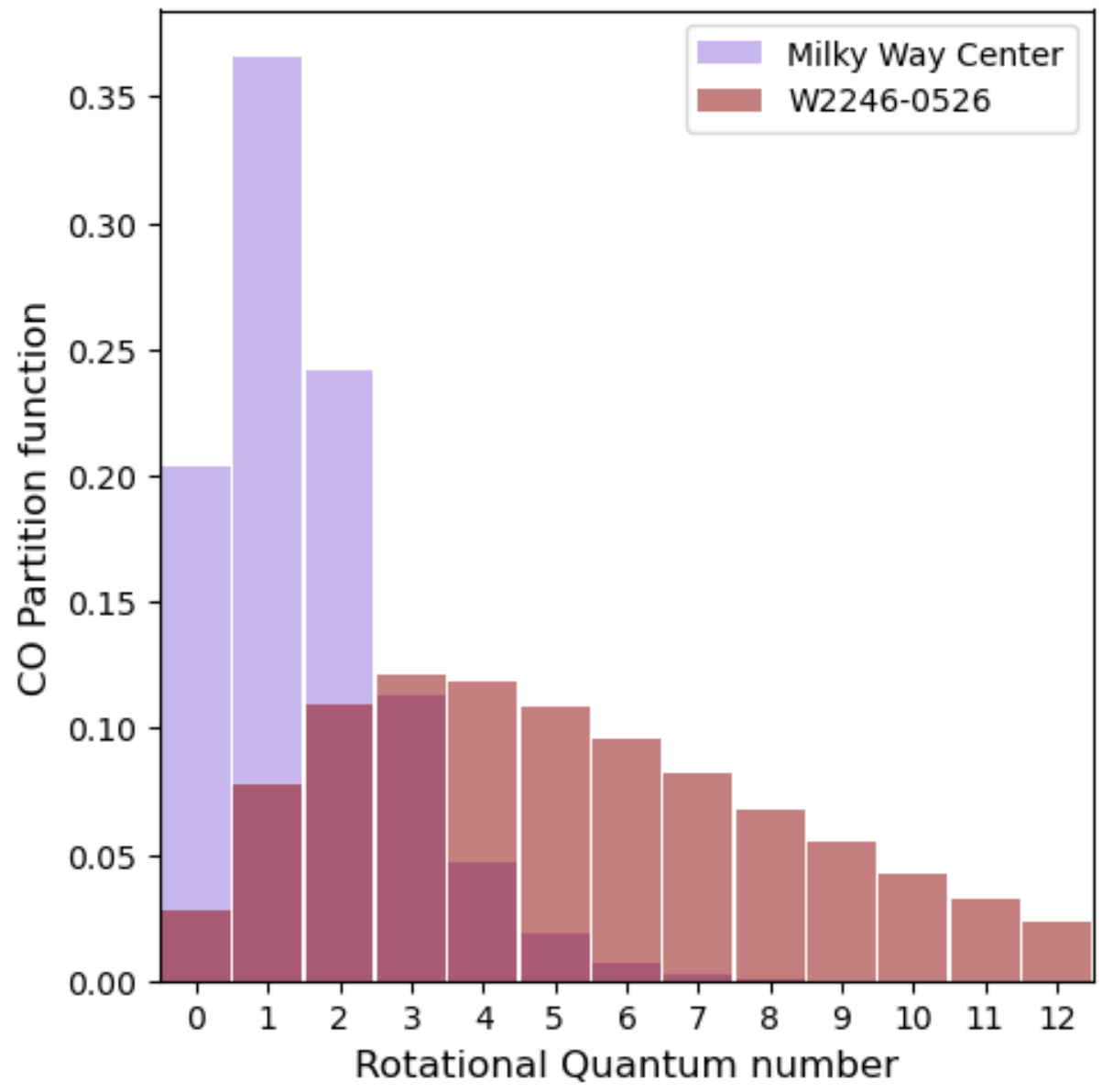

**Fig. 3.** CO partition function for W2246–0526 (red) and the Milky Way center (purple; (Fixsen et al. 1999)).

[1] Planck All-Sky Survey to Analyze Gravitationally lensed Extreme Starbursts; `https://sites.google.com/view/astropassages`

The molecular gas-to-dust-mass ratio (GDMR) and [CO/H2] gas-phase abundance are also set as free parameters, and we are able to estimate these values for the first time based on the TUNER models. As shown in Fig. B.1, the 16th-84th percentile values of [CO/H2] range between log[CO/H2] = -4.6 and -3.4, with a median value of log[CO/H2] ∼ -3.95. This is close to the local Galactic value (i.e. log[CO/H2] ∼ -4.1) at solar metallicity, yet this galaxy is observed only 1.28 Gyr after the Big Bang. Evidence still suggests a large amount of molecular gas mass, with respect to dust mass, may reside in the host of W2246–0526, with the 16th-84th percentile range of GDMR ∼ 280-690. For such a comparable CO abundance, as in local calibrations within the Milky Way, one may expect values closer to a more nominal GDMR∼150. Such values are possible (Appendix Fig.B.1), yet the mean value of GDMR∼500 suggests that, at $z$ = 4.6, this extreme object may have differences in the dust grain properties compared to the Milky Way. The inferred dust mass $M_d \sim 10^8$ $M_\odot$ implies a significant amount of dust, and therefore metals, to suggest the metallicity is likely not extremely low. Understanding the GDMR, and how the CO gas-phase abundance and other parameters may be related requires further investigation that is beyond the scope of this work. We note, however, that there is a systematic factor 2-3 in the choice of the dust mass absorption coefficient value and the frequency at which it is normalized (Draine & Li 2007; Draine 2011).

### *3.3. Understanding what CO lines trace the molecular gas mass for the highly excited W2246–0526*

Here we explore how the model-derived molecular gas mass and CO line luminosities relate to one another. This enables a self-consistent approach to better understand how some CO emission lines may better trace the molecular gas mass in W2246–0526 and similarly excited objects. That is, we explicitly calculate the CO line luminosity to molecular gas mass conversion factor $\alpha_{CO} = M_{ISM}/L'_{CO}$ [$M_\odot$ (K km s$^{-1}$ pc$^2$)$^{-1}$] and plot the normalised molecular gas mass distribution produced by the TUNER model fits (Figure 4). Then, to determine what CO lines best share the same overlap in density, the relative line luminosities are compared to the corresponding peak in the molecular gas mass distribution. This approach to determine the best CO emission lines tracing the molecular gas mass is different than what is commonly done. We are explicitly calculating the molecular gas mass via dust and CO modeling of the physical parameters without resorting to an $\alpha_{CO}$ using common assumptions. The current measurements and analyses are thereby limited by the uncertainties used to derive this conversion factor, and modeling the host within W2246–0526 implies that an even more complete picture of the total molecular gas mass content from the entire field could still be determined[2]. That is, we neglect the contribution to the total molecular gas mass from the possible extended component seen in the CO(2-1) line measurements (e.g. Fig. 1, Fig. A, as well as Díaz-Santos et al. 2018).

Following Equation 5 in Harrington et al. (2021), we calculate the mean total molecular gas mass for the central region of W2246–0526 using the model-fit parameters (and including the mass contribution from He), e.g. the emitting size radius and the central density and width of the lognormal PDF: $M_{ISM}$ = (7.75 ± 3.39) × $10^{10}$ $M_\odot$. We further derive the corresponding molecular gas mass conversion factors for the central host from the model line luminosities that best fit the observed data following (Harrington et al. 2021). Local calibrations to molecular gas mass using the CO(1-0) line are well-justified because the typical temperatures and densities are quite low, with the level populations heavily weighted by the ground-state and J=1 levels (see Figure 3). This was noted by Harrington et al. (2021) when modeling strongly lensed starbursts with $L_{IR} \sim 10^{13-14}$ $L_\odot$, revealing a wide range of observed line SEDs (and varying CO partition functions). For W2246–0526, the bulk of the CO molecules that contribute to the molecular gas mass are found to be at densities log($n$($H_2$)) [cm$^{-3}$] ∼ 3.5-5.5. The observed mid-J CO lines in the center of W2246–0526 that are responsible for the gas at these densities, e.g. J = 3 to 6, are more heavily populated than the J = 0 or 1 levels. This is contrary to low-excitation gas conditions in the Milky Way, where the bulk of the molecular gas mass can be traced by the CO(1-0) emission line. The $\alpha_{CO}$ factor for each transition can be derived for W2246–0526 using the ratio of the model-fit line luminosity for each transition and the total molecular gas mass (Harrington et al. 2021). The relative uncertainties in the model-derived conversion factor are all within 1-$\sigma$ for the low/mid-J transitions. For transitions CO(1-0), CO(2-1), CO(5-4) and CO(7-6) we find a mean value of $\alpha_{CO(1-0)}$ = 2.50 ± 2.03, $\alpha_{CO(2-1)}$ = 1.57 ± 0.9, $\alpha_{CO(5-4)}$ = 1.72 ± 0.76, $\alpha_{CO(7-6)}$ = 2.29 ± 0.98 $M_\odot$ (K km s$^{-1}$ pc$^2$)$^{-1}$. The larger uncertainty in the $\alpha_{CO(1-0)}$ value in these models largely stems from not having the measured CO(1-0) line. The extreme excitation also leads to optical depth effects resulting in solutions with optically thin CO(1-0) line emission (Harrington et al. 2021), and therefore a wider uncertainty in deriving the CO(1-0) line luminosity. Regardless, the peak of the partition function remains shifted to higher levels for this extreme galaxy, and the mass can be similarly traced by higher-J transitions. Figure 4 further demonstrates why it may be challenging to use, e.g., CO(1-0) as a tracer of the bulk molecular gas mass for such a highly excited systems such as the center of W2246–0526. In §3.1 we report that only ∼10% of the molecular gas column density is traced by the CO(1-0) line emission in this extreme source. Figure 4 shows that although the CO(1-0) line luminosity distribution versus density traces the densities which correspond to the peak of the molecular gas mass versus density distribution, the higher-J CO line luminosities have distributions centered on the mass distribution. Between the CO(7-6) to CO(12-11) line luminosity distributions there is an overlap with the peak of the molecular gas mass distribution. The resulting shift towards higher level populations (Figure 3) is a stark contrast to Milky Way conditions. For systems such as the central host of this Hot DOG, it may be easier to observe a low-to-mid-J CO line and directly apply a conversion factor of $\alpha_{CO(J=2-1\ to\ 5-4)} \sim 1.5 - 2.0$ $M_\odot$ (K km s$^{-1}$ pc$^2$)$^{-1}$. We

[2] e.g. measuring warm molecular gas within the field or faint atomic carbon line emission that may trace molecular gas in regions with photodissociated CO.

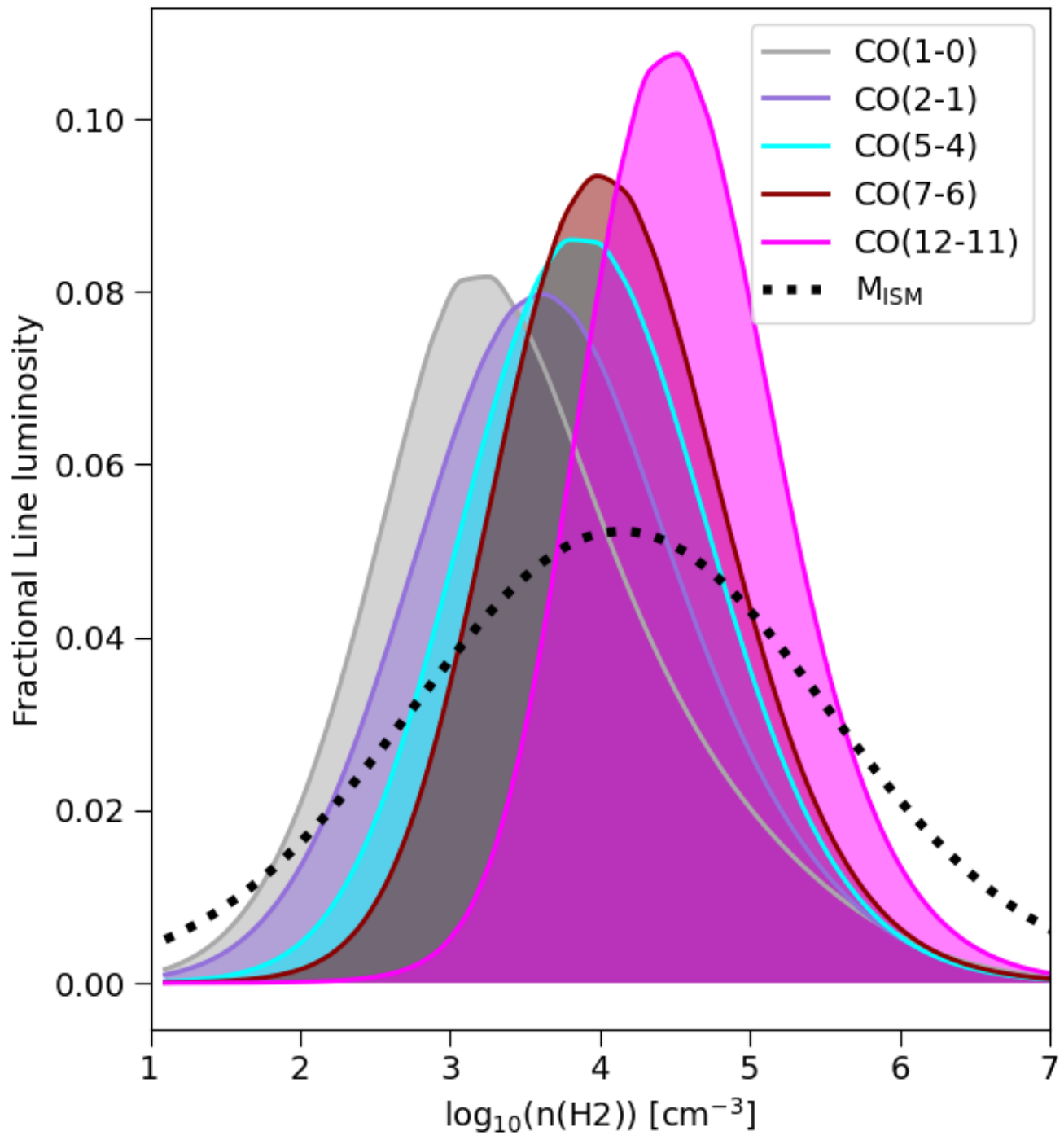

**Fig. 4.** The distribution of the fractional line luminosity per transition of CO (colored lines and shaded regions; increasing in transition from left to right) versus density, with the normalized molecular gas mass distribution (dashed black line).

note that without sensitive measurements of the mid-to-high-J CO lines in the possible extended component it is not possible to conclude that the cool gas associated to the neighboring systems (Díaz-Santos et al. 2016, 2018; Vayner et al. 2024; Fernández Aranda et al. 2025) around the center of W2246–0526 may also be better traced by the higher-J transitions, as there may be less excited CO. If the CO line SED is much less excited than the center of W2246–0526 then the lower-J lines such as CO(2-1) may better trace the environment around the host for a total molecular gas measurement, demonstrating the importance of determining the CO partition function for each system. Refined calibrations are required in a larger sample to minimize uncertainties in order to accurately determine molecular gas masses under such extreme conditions. Insights from modeling the full dust and CO line SED of W2246–0526 suggests that it may be more appropriate to measure the CO lines with J≥ 3 to trace the bulk of the molecular gas mass for such extreme systems.

## 4. Summary and Conclusions

Since 2015, W2246–0526 has been known to be among the most IR luminous galaxies ever discovered (Tsai et al. 2015). Despite this, there has yet to be a multi-J CO radiative transfer analysis of any of the *WISE*-selected Hot DOGs to characterize the physical conditions of the molecular gas and dust. There have been only a handful of QSO samples that have multiple CO lines, the most extreme example being the strongly lensed APM0827. This study therefore presents the first analysis of the rise and fall of the CO line SED of any *WISE*-selected Hot DOG, establishing a benchmark reference for future work of these extreme systems with dust-obscured nuclear activity. W2246–0526 has extreme gas excitation, with a detection of the high-J CO(12-11) emission line and evidence of the CO(17-16) line, which indicates that future high-J line observations of such high redshift objects is feasible with current facilities.

We performed a kpc-scale analyses using the TUNER model, which *simultaneously* fits both the dust and CO line SEDs in the framework of a turbulence driven lognormal probability distribution of molecular gas densities. This initial modeling of the extreme conditions indicates that that such high-J lines are required to characterize the contribution from the most dense molecular gas in this Hot DOG. Such highly excited CO line SEDs require a combination of high gas density and kinetic temperature – motivating future work to examine the rise and fall of the CO line SEDs to refine our understanding of the physical molecular gas properties in a larger sample and explore the source of such high-excitation CO (e.g. X-ray dominated regions (Gallerani et al. 2014)). The molecular gas mass close to $10^{11}$ $M_\odot$ infers an abundant amount of material within the vicinity of this active SMBH($\sim 50-100\times$ the mass of the SMBH). The plethora of sources of kinetic activity and mechanical heating have already indicated an extreme environment within W2246–0526. The highest possible angular resolution achieved by ALMA (i.e. Configuration 10) enabled a 600pc resolved [CII] study, with these high resolution results indicating a dust-obscured SMBH injecting radiative and mechanical feedback within $\sim$ 3 kpc$^2$ (Liao et al. submitted). We find independent evidence of strong mechanical heating reflected in the TUNER model results, with values of $T_k$ / $T_d$ $\sim$ 4.3 within a similar area. This ratio may be a key diagnostic for the molecular ISM excitation conditions of the most extreme systems. With current facilities it may not be feasible to further resolve all lines down to 10s to 100s of pc scales in a reasonable amount of integration time. Such efforts will require next-generation facilities to systematically examine the CO emitting regions around the core area of Hot DOGs and other systems to spatially resolve the conditions of these short-lived galaxy episodes of heavily dust-obscured SMBH activity.

*Acknowledgements.* We would like to thank the referee for their insightful and engaging comments that have significantly improved this manuscript. KCH would like to thank Amit Vishwas, Melanie Kaasinen and Carlos De Breuck for useful discussions. KCH would like to thank the European Southern Observatory for their generous science travel support during this work, and for the astronomy department at the Universidad Diego Portales for being such warm hosts last year. KCH thanks Mary and Mike Gyra for their guidance and support in the final stages of this manuscript. KCH and RFA thank the Instituto Roca Negra for their kind hospitality and work environment. MA is supported by FONDECYT grant number 1252054, and gratefully acknowledges support from ANID Basal Project FB210003 and ANID MILENIO NCN2024 112. RJA was supported by FONDECYT grant number 1231718 and by the ANID BASAL project FB210003. CWT is supported by NSFC 11988101. Part of this research was carried out at the Jet Propulsion Laboratory, California Institute of Technology, under a contract with the National Aeronautics and Space Administration (80NM0018D0004). The research leading to these results has received funding from the European Union's Horizon 2020 research and innovation programme under the Marie Skłodowska-Curie RISE action, Grant Agreement n. 873089 (ASTROSTAT-II). T.D.-S. acknowledges the research project was supported by the Hellenic Foundation for Research and Innovation (HFRI) under the "2nd Call for HFRI Research Projects to support Faculty Members & Researchers" (Project Number: 3382). This paper makes use of the following ALMA data: ADS/JAO.ALMA#2015.1.00883.S,2016.1.00668.S,2017.1.00899.S, 2018.1.00119.S, 2018.1.00333.S, 2019.1.00219.S, 2021.1.00726.S. ALMA is a partnership of ESO (representing its member states), NSF (USA) and NINS (Japan), together with NRC (Canada), NSTC and ASIAA (Taiwan), and KASI (Republic of Korea), in cooperation with the Republic of Chile. The Joint ALMA Observatory is operated by ESO, AUI/NRAO and NAOJ.

## Appendix A: VLA and ALMA multi-J CO / continuum observation and reduction details

We imaged the visibility measurement sets to produce spectral cubes for the emission lines and continuum images using the `tclean` algorithm of the Common Astronomy Software Applications (CASA; McMullin et al. 2007) v6.5 for both the VLA and ALMA datsets. We used the Very Large Array (VLA) in its B and D configuration and the Q-band receivers to observe the redshifted CO(2–1) emission line at 41.16 GHz and a bandwidth of X GHz in each sideband. The B-array observations were executed September 12, 2017 to January 16, 2018 using the same tuning and correlator setup as the previous D-array observations. The data were reduced using the VLA calibration pipeline, with additional manual flagging of poor visibilities. Details of the D-array observations are provided in Díaz-Santos et al. (2018). The data were weighted, concatenated and imaged using CASA. All images were primary beam corrected, however the primary beam is much larger than the source. We used a Briggs weighting method, which balances sensitivity and angular resolution, and cleaned down to $2\sigma$ following Fernández Aranda et al. (2024, 2025). The final cube is obtained with a Briggs Robust parameter Robust=2. This is effectively the same as a natural weighting, providing more weight to the inner antennas to achieve higher sensitivity. The final image cube has a width of 100 km $s^{-1}$.

ALMA image processing was performed in a similar manner as the VLA observations, as we clean down to $2\sigma$ with Robust=2 and 100 km $s^{-1}$ channels following the methods in Fernández Aranda et al. (2024, 2025). To generate the line cubes, we first subtracted the visibilities corresponding to the frequencies that did not contain line emission using the CASA task `uvcontsub`. For the continuum, we combined all line-free channels from each dataset. The CO(17-16) line was detected only partially at the edge of a spectral window corresponding to a separate tuning targeting the [CII]158$\mu$m emission line. We derive the estimated line flux density fitting the Gaussian amplitude while assuming a fixed FWHM of 600 km $s^{-1}$—Gaussian fits for the CO(2-1) to CO(12-11) lines give FWHM of $600^{+100}_{-50}$ km $s^{-1}$ —, and assuming the line would be located at the same redshift of 4.601 as the other CO lines. When propagating the uncertainty in the CO(17-16) Gaussian fit, we conservatively introduce amplitude and sigma errors of 10% and 20%, however for the modeling of the CO ladder and dust continuum SED we adopt a 100% uncertainty in the derived CO(17-16) line intensity. The moment-0 map (Fig. 2) was generated using only 20 channels corresponding to about one-third of the total CO(17-16) line emission. The continuum and CO line measured fluxes, the size of the synthesized beams, the depth of the observations, and the observed bands and project codes are presented in Table A.1.

To model the dust SED we adopt the extrapolated flux densities at the observed *Herschel* / PACS(160μm) and *Herschel* / SPIRE (250μm) wavelengths from the mean values and 1-sigma uncertainties reported in the resolved modified blackbody fits from (Fernández Aranda et al. 2025). This is instead of using the reported *Herschel* flux densities that include contributions from both hot/warm (mid-IR) and cold (FIR) emission from the host galaxy W2246–0526 (Tsai et al. 2018), and enables us to estimate the turnover of the dust SED with greater precision and not overestimate the far-IR luminosity. The results are consistent with known values of the derived IR luminosity from the single temperature modified blackbody fits to the core emission presented in Fernández Aranda et al. (2025).

## Appendix B: Details on the TUNER modeling

Table B lists the free parameters that are fit and the median values and their ranges between the 16th and 84th percentile values. These include the [$CO/H_2$] gas-phase abundance, the molecular gas-to-dust ratio (GDMR), mean dust temperature ($T_d$) [K], dust emissivity index ($\beta_d$), mean molecular $H_2$ gas density ($\log(n_{H_2})$ [$cm^{-3}$], mean gas kinetic temperature $T_K$ [K], turbulence dispersion velocity ($\Delta V$) [km $s^{-1}$] (i.e. the FWHM of the lognormal probablility density function distribution at a mean molecular gas density), emitting size radius ($r$) [pc] and $C_{ff}$ (unitless coefficient for the continuum foreground dimming). We have fixed the virialization parameter $\kappa_{vir}$ at 1.3, and dust mass absorption coefficient $\kappa_d = 0.047\ m^2\ kg^{-1}$ at 352.7 GHz (Draine 2011). The marginalized 1D and 2D posterior distributions for all free parameters are shown in the corner plot in Figure B.1 with the mean value and 16th and 84th percentile limits.

The lines are observed/modeled against a background continuum composed of half of the modified black-body dust continuum emission and the blackbody CMB radiation at the redshift of the object. There is a temperature floor imposed that is 10 K above the redshifted CMB radiation temperature floor in our modeling approach. Otherwise, the solutions with low temperatures would have invisible emission in contrast to the CMB radiation, leading to an inability to accurately constrain the gas excitation conditions. The temperature of the CMB radiation at the source redshift is less than 20 K. Therefore this is not a concern for W2246–0526, as both the gas and dust temperatures are about 100 K or even greater. Therefore, the temperature floor above the CMB radiation is not strictly necessary, yet it remains consistent with previous approaches in such modeling (e.g. Harrington et al. 2021). To further verify this, we also ran the model without any temperature floor. This does not change the results for any of the free parameters by more than 2%, due to the highly excited nature of this system. The only value that changes, by 20%, is the GDMR. The GDMR decreases when removing the temperature floor, but still remains within the uncertainties seen in Fig. B.1. Following Harrington et al. (2021) we constrain the parameter space to values of $T_k / T_d = 0.5 - 6.5$, which reduces degeneracies and long tails in the posterior distribution for $T_k$. We couple the $H_2$ density and gas kinetic temperature with a power-law slope index as noted in §3. Still, there can be a wide-ranging and highly degenerate parameter space. The turbulence dispersion velocity has a long tail, although the peak of the distribution is at about 150 km $s^{-1}$, which is a more plausible value. Similarly the GDMR has a broad distribution, with lower values consistent with known values of about 150-200. We employed the Markov chain Monte Carlo *emcee* Python package (Foreman-Mackey et al. 2013), with uniform priors, 100 walkers and 50 autocorrelation times, where the autocorrelation time is stable to within a few percent.

**Table A.1.** CO line and dust continuum emission measurements and details of the observations

| Continuum frequency [GHz] | Flux density [mJy] | Band | Beam [″] | Cont. depth [mJy beam$^{-1}$] | Project code |
|---|---|---|---|---|---|
| 847.0 | 51.8 ± 1.9 | ALMA B10 | 0.39 × 0.32 | 1.2 | 2017.1.00899.S |
| 607.9 | 30.0 ± 0.9 | ALMA B9 | 0.36 × 0.31 | 1.4 | 2021.1.00726.S |
| 434.1 | 12.8 ± 0.7 | ALMA B8 | 0.53 × 0.38 | 0.30 | 2016.1.00668.S |
| 360.0 | 7.45 ± 0.04 | ALMA B7 | 0.44 × 0.35 | 0.040 | 2017.1.00899.S |
| 345.3 | 7.39 ± 0.05 | ALMA B7 | 0.36 × 0.30 | 0.044 | 2018.1.00333.S |
| 253.2 | 2.66 ± 0.03 | ALMA B6 | 0.54 × 0.46 | 0.017 | 2015.1.00883.S |
| 150.1 | 0.40 ± 0.02 | ALMA B4 | 0.50 × 0.39 | 0.007 | 2019.1.00219.S |
| 93.6 | 0.071 ± 0.021 | ALMA B3 | 0.35 × 0.30 | 0.019 | 2018.1.00119.S |

| Line | Flux density [Jy km s$^{-1}$] | Band | Beam [″] | Channel depth [mJy beam$^{-1}$ per 100 km s$^{-1}$] | Project code |
|---|---|---|---|---|---|
| CO(2–1) | 0.35 ± 0.03 | VLA B+D | 0.62 × 0.46 | 0.019 | VLA/15B-192<br>VLA/17B-312 |
| CO(5–4) | 1.41 ± 0.13 | ALMA B3 | 0.38 × 0.24 | 0.51 | 2017.1.00358.S |
| CO(7–6) | 2.20 ± 0.03 | ALMA B4 | 0.50 × 0.39 | 0.35 | 2022.1.00353.S |
| CO(12–11) | 2.45 ± 0.06 | ALMA B6 | 0.56 × 0.47 | 0.84 | 2015.1.00883.S |
| CO(17–16) | 1.09 ± 1.09 | ALMA B7 | 0.35 × 0.28 | 0.40 | 2018.1.00333.S |

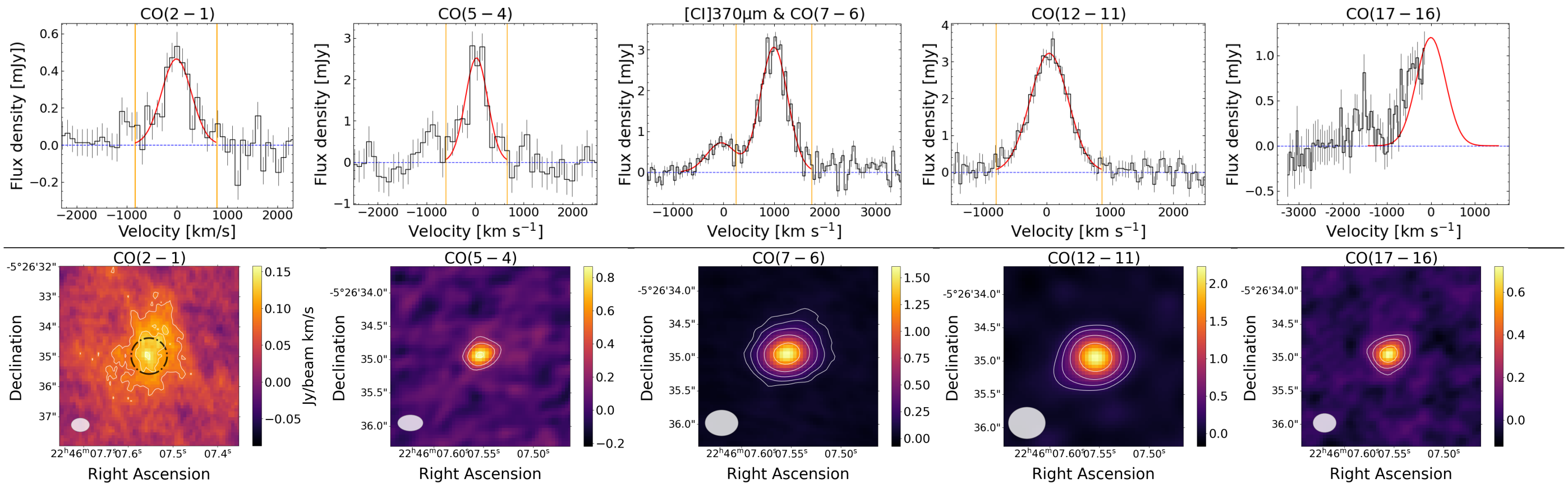

**Fig. A.1. Top:**Spectral line profiles for the measured CO lines using VLA/ALMA, centered on the same 0.6" aperture for each line, with flux density (y-axis) vs. redshifted velocity (x-axis) at the source redshift. Gaussian fits are shown in red. The CO(17-16) line is partially detected at the edge of a spectral window. **Bottom:** CO moment-0 (velocity-integrated) values in Jy/beam km s$^{-1}$ for the measured CO(2-1), CO(5-4), CO(7-6), CO(12-11) and CO(17-16). The black-dashed line in the CO(2-1) moment-0 image indicates the $r = 0.6''$ aperture used for extraction for all emission lines, corresponding to the central host in W2246–0526. We note that the map for CO(17-16) is generated using the available channels covering the line emission and is not corrected for the missing flux.

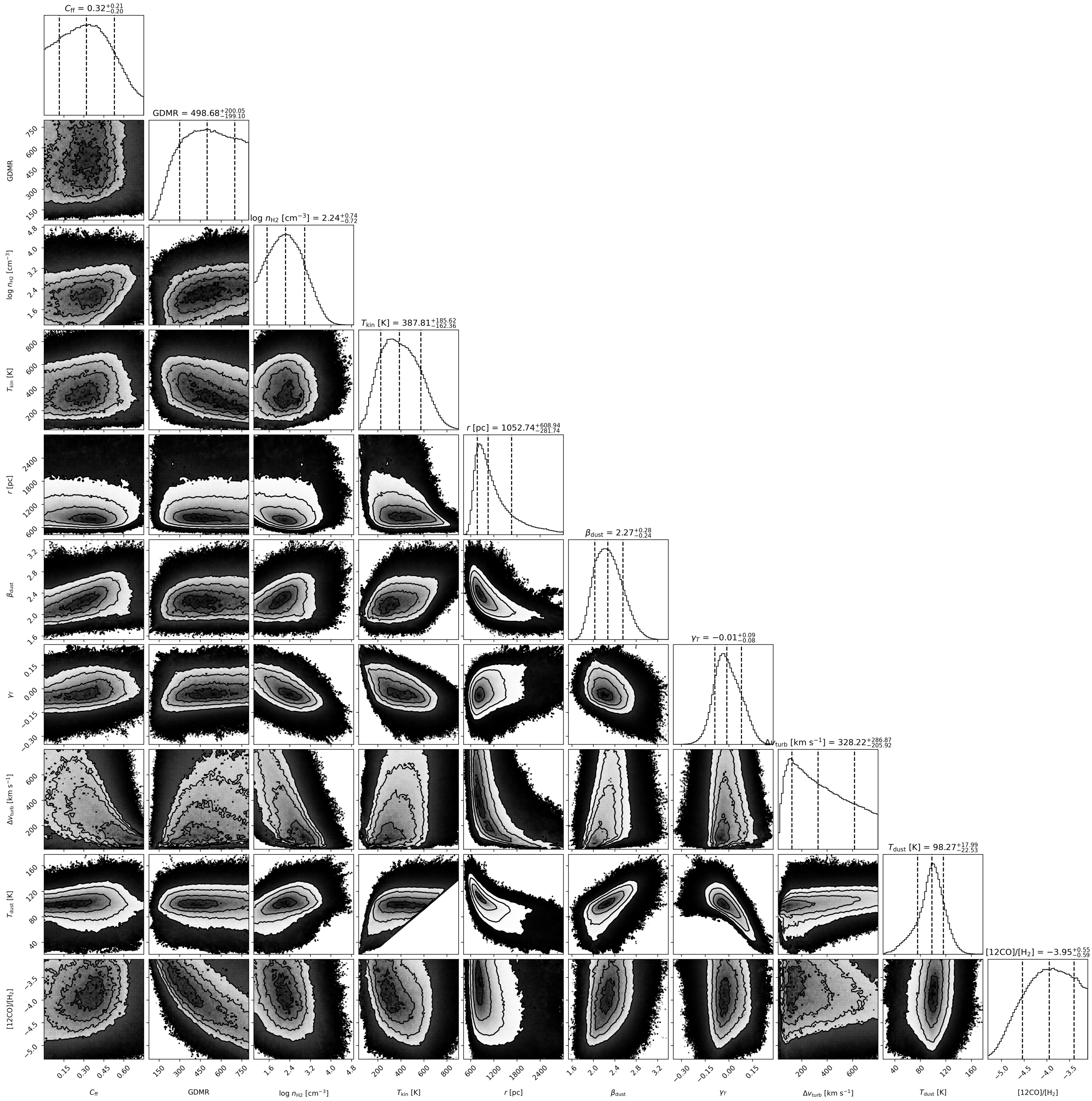

**Fig. B.1.** Corner plot of the marginalized 1D and 2D posterior distributions of the free parameters optimized in our TUNER model fits. Dashed lines indicate the mean value and 16th and 84th percentile range for each parameter (from top to bottom and left to right: $C_{ff}$, molecular ISM gas to dust ratio, GDMR, molecular $H_2$ gas density $\log(n_{H_2})$ [cm$^{-3}$], gas kinetic temperature $T_K$ [K], emitting size radius ($r$) [pc], dust emissivity index ($\beta_d$), $\gamma_T$, turbulence dispersion velocity ($\Delta V$) [km s$^{-1}$], dust temperature $T_d$ [K] and [CO/$H_2$] gas-phase abundance).

| Parameter | Value |
|---|---|
| **Physical parameters** | |
| Radius [kpc] | $1.10^{+0.56}_{-0.33}$ |
| $\log_{10}(n_{H_2})$ [cm$^{-3}$] | $2.19^{+0.72}_{-0.71}$ |
| $T_{kin}$ [K] | $405.09^{+150.56}_{-149.30}$ |
| $\Delta v_{turb}$ [km s$^{-1}$] | $296.78^{+339.38}_{-180.92}$ |
| $\kappa_{vir}$ | 1.3 (fixed) |
| $\gamma_T$ | $-0.01^{+0.09}_{-0.08}$ |
| $C_{ff}$ | $0.35^{+0.22}_{-0.25}$ |
| **Chemical parameters** | |
| $^{12}CO/H_2$ | $-3.98^{+0.54}_{-0.58}$ |
| GDMR | $505.25^{+186.03}_{-224.46}$ |
| $\kappa_d$ [m$^2$ kg$^{-1}$] | 0.047 (fixed) |
| $\nu_0, \kappa_d$ | 352.7 GHz (fixed) |
| **Radiation field** | |
| $T_{dust}$ [K] | $97.17^{+20.21}_{-24.23}$ |
| $\beta_{dust}$ | $2.29^{+0.30}_{-0.26}$ |
| z | 4.601 (fixed) |

**Table B.1.** The 50th percentile parameter values and 16th and 84th percentile upper and lower limits. Input model parameters kept fixed are indicated.